\documentclass[aps,pra,amsmath,amssymb,twocolumn,showpacs,
nofootinbib 
]{revtex4-1}
\pdfoutput=1
\usepackage{epsfig}
\usepackage{subfigure}
\usepackage{amsmath, amssymb, amsthm, amsfonts, siunitx, youngtab, ytableau, color, esvect}
\usepackage{array}
\usepackage{bm,mathtools,relsize}
\usepackage{braket}
\usepackage[english]{babel}
\usepackage[normalem]{ulem}
\usepackage{float}
\usepackage{graphicx,subfigure,natbib,multirow,float,dsfont}
\usepackage{hyperref}

\begin{document}
\title{On unitary reconstruction of linear optical networks}
\author{
	Max Tillmann$^1$,
	Christian Schmidt$^{1}$,
		Philip Walther$^{1}$}
\affiliation{$^1$Faculty of Physics, University of Vienna, Boltzmanngasse 5, A-1090 Vienna, Austria}

\begin{abstract}
\noindent 
Linear optical elements are pivotal instruments in the manipulation of classical and quantum states of light. Recent progress in integrated quantum photonic technology enabled the implementation of large numbers of such elements on chip, in particular passively stable interferometers. However, it is a challenge to characterize the optical transformation of such a device as the individual optical elements are not directly accessible. Thus only an effective overall transformation can be recovered. Here we present a reconstruction approach based on a global optimization of element parameters and compare it to two prominently used approaches. We numerically evaluate their performance for networks up to $14$ modes and various levels of error on the primary data.
\end{abstract}

\maketitle

\section{Introduction}
Linear optical quantum computing has attracted major attention since Knill, Laflamme and Milburn have introduced a scheme for efficient quantum computation in 2001~\cite{Knill2001}. Notwithstanding tremendous technological progress, the experimental realization of such computers is still challenging with steep requirements for the generation, manipulation and detection of the quantum states of light. In the last decade, integrated quantum photonics~\cite{Politi2009} has become central to the technological progress by providing the means to miniaturize and mass fabricate vital components, such as quantum light sources~\cite{Minkov2016,Zhang2016,Silverstone2015,Harris2014}, quantum storage devices~\cite{Konoike2016,Corrielli2016} and highly efficient photon detection on chip~\cite{Najafi2015,Kahl2015,Akhlaghi2015,Pernice2012}. Of particular importance is the manipulation of the quantum states of light via linear optical networks (LONs). Here, integrated quantum photonics enabled the fabrication of LONs with unprecedented levels of interferometric complexity. The inventory includes optical elements, facilitating either fixed ~\cite{Politi2008,Marshall2009,Crespi2011,Heilmann2014,Corrielli2014} or tunable~\cite{Thompson2011,Shadbolt2012,Metcalf2014,Carolan2015} single-qubit transformations but also novel hybride elements~\cite{Lebugle2015}. Arranging several of these elements allows fabrication of miniaturized versions of logic gates which are essential in quantum information processing. Those gates resemble small to medium scale interferometers and are either tailored to perform a particular task or can be reconfigured for a variety of tasks~\cite{Politi2008,Crespi2011,Metcalf2014,Carolan2015}. Here, the physical encoding scheme, e.g. a dual-rail polarization encoding, determines how these gates, unitary matrices acting on logical qubits, must be compiled from different optical elements. Hence, the overall transformations of these LONs are required to be unitary, too. This poses a challenge for experimental realizations which are inevitably afflicted by imperfections. Here, the major contributors are on one hand losses, which render these devices non-norm preserving. On the other hand fabrication imperfections stemming from the production itself cause deviations between an initially targeted transformation and the implemented one. Obtaining precise knowledge of an optical transformation at hand therefore is important and serves multiple purposes: It allows theoretical modelling of an experiment and allocation of the error budget to different sources of error. More important, identifying deviations or even defective optical elements is essential for troubleshooting experiments and improving future versions of an optical circuit. The ability to construct high-fidelity gate operations becomes a stringent requirement when circuits are scaled up beyond proof of principle implementations~\cite{Martinis2015}. Since all waveguides are embedded into a bulk material, only the input and output ports and therefore the overall transformation is directly accessible. To acquire knowledge of such an overall transformation one can use different approaches relying on quantum resources~\cite{Childs2001,Mitchell2003,OBrien2004,Peruzzo2011a,Laing2012} or classical resources~\cite{Lobino2008,Rahimi-Keshari2011,Rahimi-Keshari2013,Fedorov2015}. The majority of these techniques fall in the category of quantum process tomography ('QPT'). While QPT is well established in quantum science, it faces challenges when applied to large networks due to quickly growing resource costs. Alternative techniques, reconstructing a transition matrix~\cite{Rahimi-Keshari2013} or a unitary matrix description~\cite{Laing2012} of a LON, became prominent in the context of BosonSampling~\cite{aaronson2011computational,Broome2013,Spring2013,Tillmann2013,Crespi2013}. These methods scale more favourably with respect to the required measurements when compared to QPT, and reconstruct matrix descriptions of LONs omitting global phases at the input and output ports. Methods which rely on light that is scattered out of the LON are not further considered as the technique relies on loss that compromises the guiding properties of the whole structure~\cite{Spring2013}.\\
Here, we present a new approach to characterize the transformation implemented by LONs. We choose to enforce unitarity from the start by parametrizing the LONs as interferometers composed of beam splitters and phase shifters~\cite{Reck1994}. The reconstructed unitary matrices are then obtained by optimizing the beam splitting ratios and phase shifts to best explain a set of data sensed via probe states injected into the network. We utilize a data set composed of two-photon interference visibilities~\cite{Hong1987}, rendering the procedure insensitive to input and output loss. In this way both afore mentioned purposes of network reconstruction are fulfilled simultaneously; generating a description to model experimental data and gathering knowledge about the transformation of individual optical elements. Our method is a departure from the strategy of related approaches~\cite{Peruzzo2011a,Rahimi-Keshari2013}, which aim to reconstruct transition matrices, or do so in a first step~\cite{Laing2012,Dhand2016}. Whereas transition matrices are sufficient to characterize a LON and to model experimental results, they do not allow to identify deviations of individual elements directly. Here a indirect route via a polar decomposition~\cite{Laing2012,Dhand2016} and subsequent decomposition into individual elements must be chosen. The identification of faulty elements through a decomposition procedure requires LONs exhibiting a non-redundant layout, e.g. Reck et al. type networks.\\

\section{Different approaches to reconstruction}\label{sec:diff_approaches}

In the following we will compare three approaches to LON characterization. 'Brisbane'~\cite{Rahimi-Keshari2013} and 'Bristol'~\cite{Laing2012}\footnote{During the course of this work we learned of related work~\cite{Dhand2016}, which builds on~\cite{Laing2012}. Due to runtime constraints it could not be included in the numerical evaluation.} were chosen for their frequent usage in experiments relying on integrated LONs. Our approach, subsequently labelled 'Vienna', is formalizing ideas developed in~\cite{Tillmann2013,Tillmann2015}. Similar to~\cite{Peruzzo2011,Dhand2016}, an over-complete set of primary data can be used to increase reconstruction fidelities, although we find the effect to be minor (see figure~\ref{results:comparison}). In the following, primary data is referring to data sensed for reconstruction purposes in any of the approaches.\\
The three compared approaches differ in their strategy to characterize a LON. 'Bristol' and 'Brisbane' first reconstruct the individual matrix entries of a scattering description and then require a polar decomposition step to recover the closest unitary matrix. Both utilize a sufficient set of data for this purpose and in the following we refer to this strategy as a passive approach. In contrast 'Vienna' follows an active approach utilizing a larger set of primary data in a global optimization routine which already implements the unitary constrains from the beginning.\\
The approach 'Brisbane' aims to reconstruct a description of a black-box linear optical network from a sufficient set of primary data generated with coherent probe states. This data is then mapped one to one unto a scattering matrix representation $\mathbf{M}$ without the need to apply any further algorithms. The scattering matrix $\mathbf{M}$ represents a submatrix of a larger unitary matrix $\mathbf{U}$, where the additional modes of $\mathbf{U}$ correspond to loss modes. Hence, the task of loss modelling translates to the task of finding a loss matrix that couples the interferometer modes of $\mathbf{M}$ to the loss modes of $\mathbf{U}$. This necessarily includes input and output loss terms, as the approach processes transition amplitude data. 'Brisbane' covers the case of mode-dependent input loss in the following way: the loss term for each input mode $k$ is directly given by the ratio between the total power exiting the LON and the power injected into mode $k$. Subsequently the input loss is modelled by virtual beam splitters, where the square root of the loss terms corresponds to the transmittivity of the virtual beam splitters. Note that such a loss modelling works only in the case when the mode-dependent output loss of the network is zero. In general, loss inside a network cannot be parametrized this way and thus it remains unclear how a more evolved loss modelling can be included in 'Brisbane'. Hence, only a partially loss modelling scattering description $\mathbf{M}'$, which is closer to the unitary description $\mathbf{U}$ than the matrix $\mathbf{M}$, can be found via this method.\\
Experimental environments exhibiting loss as detailed above cause $\mathbf{M}'$ to be noticeably non-unitary and necessitate a polar decomposition if the closest unitary description $\mathbf{\tilde{U}}$ is to be obtained. This polar decomposition can introduce further error dependent on the size of the interferometer under test~\cite{Edelman1988}.\\
Up to the deviations introduced by the polar decomposition the entries of $\mathbf{\tilde{U}}$ reconstructed via this procedure are identical to the primary data generated. This self-consistency proves to be challenging when assessing the fidelity and the uncertainty of a reconstructed matrix in the presence of measurement errors. The complete set of single-input data and phase data is already used to fix the real entries and phases of $\mathbf{M}$. To obtain realistic error estimations for the individual entries of $\mathbf{M}$ or $\mathbf{\tilde{U}}$, the various error sources need to be studied in detail, including calibration uncertainty of the detector efficiencies and uncertainties introduced by fiber mating and coupling to a waveguide. Opposed to more complex algorithmic approaches like 'Bristol' or 'Vienna', the sensed data cannot be used to generate a quantifier for reconstruction success. An additional set of data generated by different means than coherent states is required for this purpose, e.g. a set of two-photon interference visibilities as done in~\cite{Rahimi-Keshari2013}.\\
The approach 'Bristol' aims to reconstruct a unitary description of a black-box linear optical network from a set of primary data, which in this case is generated via quantum probe states. The magnitude of each phase is sensed via a visibility of a non-classical two-photon interference but the sign of the phase needs to be calculated in relation to the other phases such that unitary constraints are obeyed. Note that a phase sensed this way is not a direct phase measurement like in the approach 'Brisbane'. Furthermore the visibility of each non-classical interference is also modified by the four contributing real entries $\tau_{jk}$ of the scattering matrix, with $j$ and $k$ labelling the output and input modes, respectively. These transition amplitudes are calculated in a fashion insensitive to mode-dependent input and output loss: the loss terms drop out by relating all input and output single-photon count rates to each other. Therefore each entry of the reconstructed matrix, $\mathbf{M_{jk}=\tau_{jk}e^{i \theta_{jk}}}$, becomes dependent on the whole set of primary data. All $\tau_{jk}$ and $\theta_{jk}$ are recovered by solving a linear system of non-linear equations. Again the closest unitary matrix, $\mathbf{\tilde{U}}$, can be found by applying a polar decomposition. Whereas the method is insensitive to loss at the input and output ports of a LON, it is sensitive to mode-dependent propagation loss. The latter introduces systematic error in the algorithm, already before applying the polar decomposition (Further detail on the influence of mode-dependent propagation loss for all three reconstruction approaches is given in Appendix D). Ultimately, the combination of the non-linear dependencies in the algorithmic approach with just a sufficient set of data leads to a lower performance of 'Bristol' with respect to 'Brisbane' and 'Vienna'.\\

\begin{figure}[H]
\centering
\includegraphics[width=0.48\textwidth]{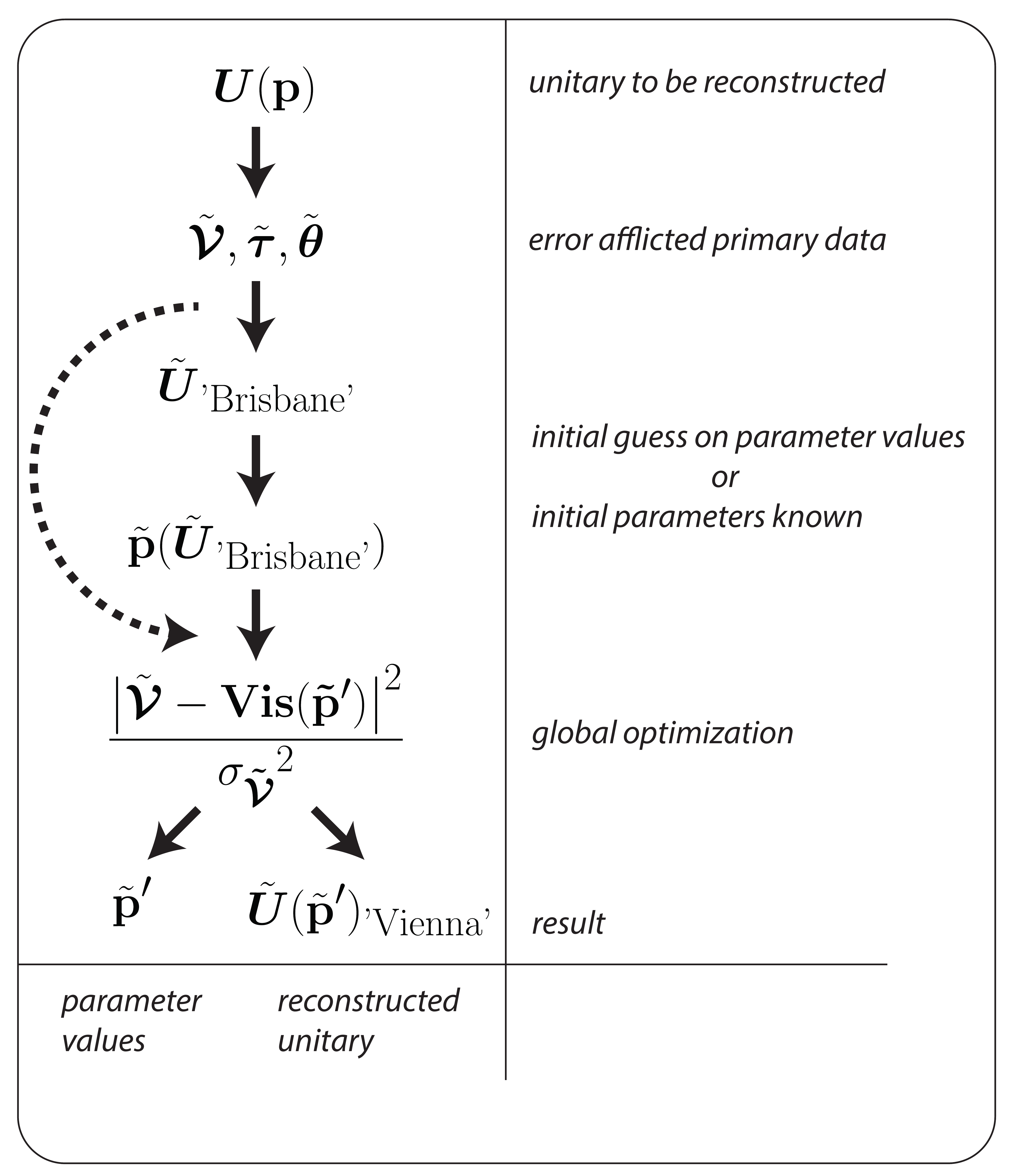}
\caption{\textbf{Flowchart for 'Vienna'.} Primary data $\tilde{\boldsymbol{\mathcal{V}}},\ \tilde{\boldsymbol{\tau}},\ \tilde{\boldsymbol{\mathcal{\theta}}}$ from a LON described by the unitary matrix $\boldsymbol{U}(\boldsymbol{p})$ is measured. In general this data will be error afflicted denoted by the tilde. The layout and initial parameters $\boldsymbol{p}$ are either known from fabrication (dashed arrow) or are obtained by a reconstruction step, e.g. 'Brisbane'. These initial parameters are now subjected to a global optimization using an, in the best case over-complete, set of primary data. Here the output yields both, the reconstructed unitary $\tilde{\boldsymbol{U}}(\tilde{\boldsymbol{p}}')_{\text{'Vienna'}}$ and the parameters of the individual building blocks $\tilde{\boldsymbol{p}}'$.} 
\label{methods:flowchart}
\end{figure}

Our new approach 'Vienna' aims to reconstruct the unitary matrix descriptions $\boldsymbol{U}(\boldsymbol{p})$ of a LON via a global optimization of optical element parameters, $\boldsymbol{p}$, which is visualized in figure~\ref{methods:flowchart} as a flowchart. For purpose-built networks, e.g. quantum logic gates, the physical layout of the optical elements and their target parameters are defined by the encoding scheme and type of gate. If they are sufficiently precise, these target parameters can serve as initial guesses for $\boldsymbol{p}$ and therefore as the starting parameters in the optimization routine. Black-box $m \times m$ LONs can be represented by an arrangement of $n=\binom{m}{2}$ beam splitters and $n-1$ phase shifters~\cite{Reck1994} (see the Appendix for a sketch). In our approach it is sufficient to obtain just one representation of the physical network decomposed in such a way. However, without any knowledge of the starting values the minimization of the $\text{p}$-dimensional landscapes is prone to converge into some local minimum only. Hence the starting values need to be obtained by different means. Here we utilize one of the passive reconstruction approaches and find that the approach 'Brisbane' is better suited for this purpose than the approach 'Bristol'. Note that both approaches lead to reconstructed unitaries that are equivalent to the unitary decomposed via the Reck et al. scheme modulo diagonal phase matrices. These diagonal phase matrices do not affect the extraction of the starting parameters~\cite{Peres1989}. To reconstruct $\mathbf{U}(\mathbf{p})$ a set of primary data, $\tilde{\boldsymbol{\mathcal{V}}},\ \tilde{\boldsymbol{\tau}},\ \tilde{\boldsymbol{\mathcal{\theta}}}$, is recorded, where $\tilde{\boldsymbol{\mathcal{V}}}$ denotes the full set of two-photon interference visibilities and $\tilde{\boldsymbol{\tau}}$ and $\tilde{\boldsymbol{\mathcal{\theta}}}$ denote the full set of normalized transition amplitudes and relative phases sensed via coherent states, respectively\footnote{Throughout the numerical evaluation the transition amplitude data is normalized such that $\sum\limits_{j=1}^m|\tau_{jk}|^2=1$.}. This data will be in general error afflicted indicated by the tilde. $\tilde{\boldsymbol{\tau}},\ \tilde{\boldsymbol{\mathcal{\theta}}}$ are only used in the case of black-box networks to obtain the initial starting parameters for the global optimization. Finally, a global cost function using an over-complete set of two-photon interference visibilities, $\tilde{\boldsymbol{\mathcal{V}}}$, is minimized to obtain optimal reconstructed parameters $\tilde{\boldsymbol{p}}'$. These automatically yield the reconstructed unitary $\tilde{\boldsymbol{U}}(\tilde{\boldsymbol{p}}')_{\text{'Vienna'}}$. Optimizing a global cost function comes with an additional advantage: the minimum of that function can act as an direct estimator for the success of the reconstruction and is in our case identical to the $\chi^2$~\cite{taylor1997introduction}, allowing further statistical interpretation. We choose to utilize just two-photon interference visibilities for the reconstruction of $\tilde{\boldsymbol{U}}(\tilde{\boldsymbol{p}}')_{\text{'Vienna'}}$. These visibilities are insensitive to input and output loss, thus $\tilde{\boldsymbol{U}}(\tilde{\boldsymbol{p}}')_{\text{'Vienna'}}$ represents a unitary description modulo loss matrices at the input and output. The parameters of these loss matrices can be easily recovered by using loss sensitive data such as transition amplitude data, $\tilde{\boldsymbol{\tau}}$, and solving a system of linear equations utilizing the reconstructed description $\tilde{\boldsymbol{U}}(\tilde{\boldsymbol{p}}')_{\text{'Vienna'}}$.

\section{Results}\label{sec:results}

\begin{figure}[b!]
\centering
\includegraphics[width=0.53\textwidth]{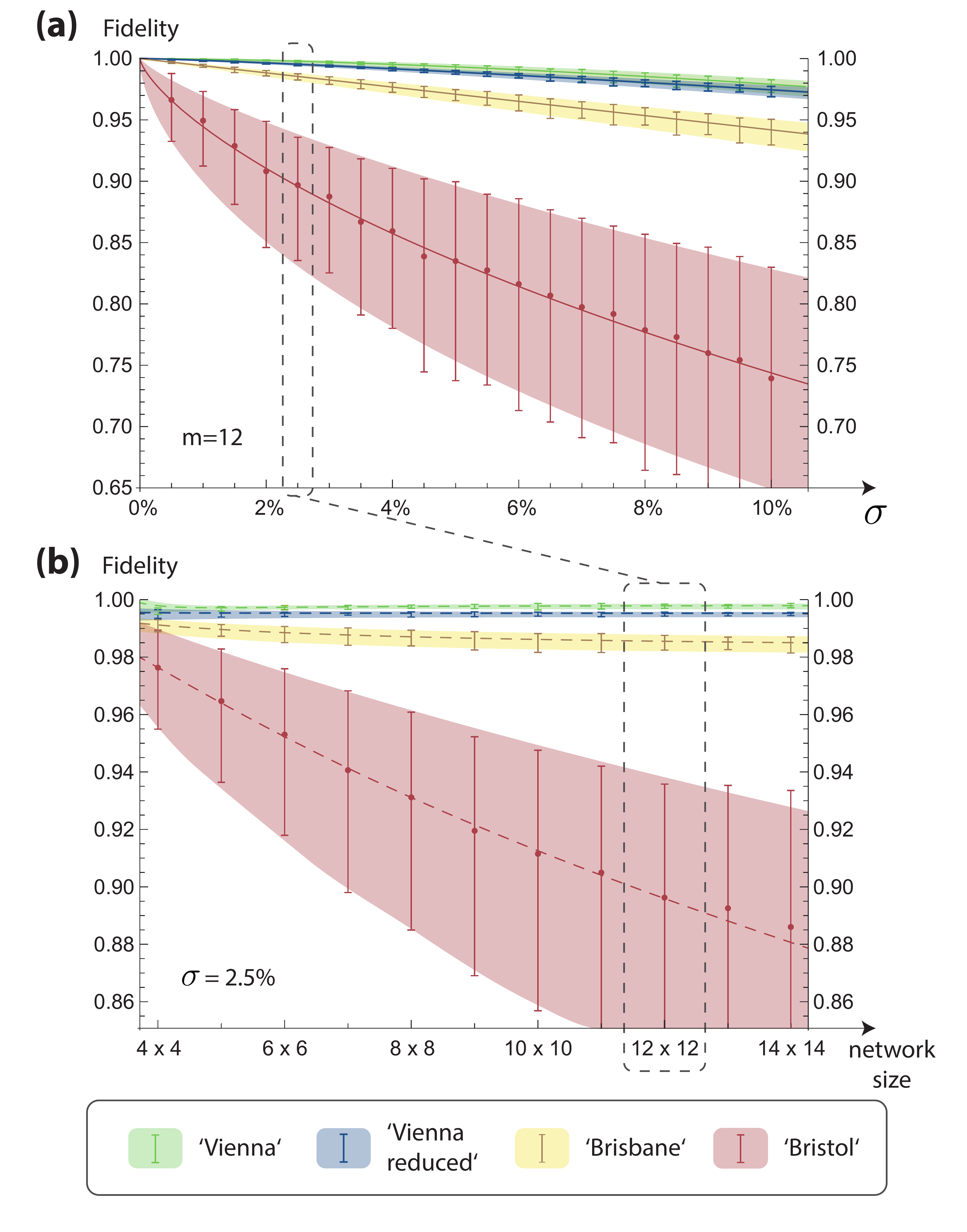}
\caption{\textbf{Comparison of the reconstruction performance} obtained with the different approaches, 'Vienna', 'Brisbane', and 'Bristol'. 'Vienna reduced' denotes a variant of 'Vienna' utilizing a smaller set of primary data (see Appendix C for further information). \textbf{(a)} shows how the fidelities and their uncertainties scale dependent on the error of the primary data, $\sigma$ for the case of $12\times12$ networks, whereas \textbf{(b)} shows the scaling dependent on the size of the $m\times m$ networks for $\sigma=2.5\%$. \textbf{(a)} and \textbf{(b)} intersect, indicated by the dashed grey boxes.} 
\label{results:comparison}
\end{figure}

We compare the reconstruction results for the different approaches 'Brisbane', 'Bristol' and 'Vienna' numerically for fully coupled $m \times m$ networks in the presence of perturbance on the primary data. To quantify the performance of the different approaches the fidelity between the initially generated Haar-random unitary matrix, $\mathbf{H}_{m,j}$, and the reconstructed unitary matrix $\boldsymbol{\tilde{U}}_{m,j,\sigma,\mu}$ is calculated. We consider $m\times m$ networks with $m=4,\,\dots,\,14$, where $\mu$ labels the reconstruction approach and $\sigma$ denotes the level of perturbance on the primary data (see Appendix A for further details). Losses are kept zero to allow for a fair comparison between loss sensitive and insensitive approaches. For each network size, $120$ Haar-random unitary matrices are generated (labelled by $j$)\footnote{For 'Bristol' always $10^3$ matrices are sampled due to the dispersed results.} to ensure that random properties of a $j$th unitary, e.g. symmetry, do not lead to biased results. The unitary descriptions are calculated via a Monte Carlo method drawing the data required for each reconstruction approach, $\mu$, randomly from the set of perturbed data, $(\boldsymbol{\tilde{\mathcal{V}}},\boldsymbol{\tilde{\tau}},\boldsymbol{\tilde{\theta}})_{m,j,\sigma}$. An average unitary description, $\boldsymbol{\tilde{U}}_{m,j,\sigma,\mu}$, is obtained after $120$ iterations with $\sigma(\boldsymbol{\tilde{U}}_{m,j,\sigma,\mu})$ denoting its standard deviation. We use the fidelity measure,

\begin{equation}\label{eq:fid}
\mathbf{F}_{m,j,\sigma,\mu}\left(\mathbf{H}_{m,j}, \boldsymbol{\tilde{U}}_{m,j,\sigma,\mu}\right )=
1-\frac{\left\Vert \mathbf{H}_{m,j}-\boldsymbol{\tilde{U}}_{m,j,\sigma,\mu} \right\Vert}{2m}
\,, 
\end{equation}

which is normalized by the number of modes, $m$, such that it is insensitive to the network size. Here, $\left\Vert.\right\Vert$ denotes the trace norm. For given $m,\, \sigma,\, \mu$ the resulting fidelity histograms are fitted with Weibull distributions centred around the most probable value, $\mathbf{\tilde{F}}_{m,\sigma,\mu}$ (see Appendix A). As an error measure, $\sigma_{\frac{1}{e}}(\mathbf{\tilde{F}}_{m,\sigma,\mu})$, the distances between the most probable fidelity and the two fidelities where the maximum probability decreased to $\frac{1}{e}$ are used. The most probable fidelities and their respective uncertainties for $200$ different combinations of network size and perturbance on the primary data are recovered for each of the reconstruction approaches. Figure~\ref{results:comparison} shows a representative example, once for $12\times12$ networks and variable error on the primary data, $\sigma$, and once for an error of $\sigma=2.5\%$ and variable network size. Clear differences between the approaches can be observed with respect to the overall performance, the scaling and the dispersive behaviour. Such differences must be attributed to specifics of the reconstruction algorithms as all approaches reconstruct the same unitaries $\mathbf{H}_{m,j}$.

The approach 'Brisbane' shows high reconstruction fidelities with low dispersion which scale linearly with the error, $\sigma$, on the primary data (figure \ref{results:comparison}\textbf{(a)}). Here, an upper bound to the deviation of the reconstructed unitaries, $\boldsymbol{\tilde{U}}_{m,j,\sigma}$, from the initial one, $\mathbf{H}_{m,j}$, in the Frobenius norm can be even given analytically:
\begin{equation}
\left\Vert \tilde{\boldsymbol{U}}-\boldsymbol{H}\right\Vert\leq(1+\sqrt{2})\kappa(\boldsymbol{M})\cdot\epsilon+\mathcal{O}(\epsilon^{2}).\label{eq: Error Brisbane}
\end{equation}
Where $\boldsymbol{M}$ and $\kappa(\boldsymbol{M})=\left\Vert \boldsymbol{M}^{-1}\right\Vert \cdot\left\Vert \mathbf{M}\right\Vert$ denote the reconstructed matrix before applying a polar decomposition and its condition number, respectively. Hence the deviation, $\epsilon=\left\Vert \mathbf{M}-\mathbf{H}\right\Vert$, stems from the polar decomposition, which in turn is due to the errors of the primary data that contribute in first order approximation as $M_{ij}\rightarrow H_{ij}+\delta H_{ij}$,
where $\delta H_{ij}=H_{ij}(i\delta\theta_{ij}+\frac{\delta \tau_{ij}}{\tau_{ij}})$. The data shown in figure \ref{results:comparison}\textbf{(b)} indicates, that the performance of the approach 'Brisbane' is only slightly affected by the network size.\\
The reconstruction fidelities yielded with the approach 'Bristol' are dominated by a (sub)exponential decay, both as a function of the error on the primary data and the network size. This also leads to a highly dispersive behaviour which is reflected in the comparably large error bars in figure \ref{results:comparison}. While the exponential decay is already observed in the original publication~\cite{Laing2012}, this just represents a phenomenological fit to the data. We conjecture that on one hand this scaling originates from the additional matrix inversion that has to be taken into account when the transition amplitudes are calculated. On the other hand, the primary data undergoes a more complex algebraic transformation, which, dependent on the noise of the primary data, can cause unfavourable amplification of perturbations.\\
Reconstructing the unitary description of unknown $m\times m$ networks via the approach 'Vienna' or 'Vienna reduced' works with highest fidelity and minimal uncertainty. Here, 'Vienna reduced' denotes a variant of 'Vienna' utilizing a smaller set of primary data (see Appendix C for further information). This behaviour can be primarily attributed to the natural implementation of the unitary constraints in the algorithm. The statistical advantage of a full over-complete set of primary data as used in 'Vienna' over a smaller set of primary data as used in 'Vienna reduced' is noticeable, albeit being small. For both approaches we find that errors on the starting parameters are of greater impact than errors on the primary data sets of two-photon visibilities. This was tested via a separate numeric evaluation. All starting parameters are extracted using the approach 'Brisbane' and as a consequence the scaling with the error on the primary data, $\sigma$, is inherited for large $\sigma$. A better scaling is found for small $\sigma$, as more precise starting parameters increase the chance that the optimization routine will converge into the global minimum. Both, 'Vienna' and 'Vienna reduced', exhibit negligible dependence on the size of the $m\times m$ networks.

\section{Discussion}\label{sec:discussion}
Precise knowledge about the optical transformation of LONs is a requirement for the validation of experimental results against theoretical predictions in numerous experiments~\cite{Peruzzo2010,Sansoni2012,Metcalf2014,Biggerstaff2016}. For this purpose, the optical transformations can be given either in terms of scattering descriptions or unitary transformations, where the latter allow a decomposition into individual building blocks. Hence the element parameters for each optical element in the LON can be obtained. From a technological point of view this is beneficial as it enables the localization of erroneous elements. Here we present a new approach, 'Vienna', and compare it to two prominently used approaches, 'Brisbane' and 'Bristol'. We investigate all approaches for the regime of zero mode-dependent loss and quantify the differences in reconstruction performance of unitary descriptions via an extensive numerical evaluation: more than $10^5$ $m\times m$ black-box networks are sampled for distinct $m$ from primary data exhibiting various levels of perturbance. The results substantiate that the direct implementation of the unitary constraints, as done in the approach 'Vienna', are of advantage for highest reconstruction fidelities. Two-photon interference visibilities play a unique role as they are a priori insensitive to input and output loss and allow to obtain over-complete sets of primary data, which are beneficial for highest reconstruction precision.\\
\newpage

\newpage

\begin{acknowledgments}
We acknowledge Valentin Stauber and Frank Verstraete for technical assistance with the numerical evaluation. The authors thank Borivoje Dakic and Si-Hui Tan for helpful discussions.\\
C.S. acknowledges support from a DAAD Jahresstipendium. M.T., C.S., and P.W. acknowledge support from the European Commission with the project EQuaM - Emulators of Quantum Frustrated Magnetism (No. 323714), GRASP -Graphene-Based Single-Photon Nonlinear Optical Devices (No. 613024), PICQUE - Photonic Integrated Compound Quantum Encoding (No. 608062),  and QUCHIP - Quantum Simulation on a Photonic Chip (No. 641039), the Vienna Center for Quantum Science and Technology (VCQ), and the Austrian Science Fund (FWF)
with the projects PhoQuSi Photonic Quantum Simulators (Y585-N20) and the doctoral programme CoQuS Complex Quantum Systems, the Vienna Science and Technology Fund (WWTF) under Grant No. ICT12-041, and the Air Force Office of Scientific Research, Air Force Material Command, United States Air Force, under Grant No. FA9550-1-6-1-0004.\\
The authors declare that they have no competing financial interests.\\
\end{acknowledgments}

\clearpage

\appendix

\begin{widetext}
\section*{APPENDIX A: Numerical evaluation of the different approaches in the lossless case}\label{sec:methods}

To quantify the performance of the different reconstruction approaches a fidelity between the initially generated Haar-random unitary matrix, $\mathbf{H}_{m,j}$, and the reconstructed unitary matrix $\boldsymbol{\tilde{U}}_{m,j,\sigma,\mu}$ is calculated. Here $\mu$ denotes the reconstruction approach and $\sigma$ the level of perturbance on the primary data. For each network size, $120$ Haar-random unitary matrices are generated (labelled by $j$) to ensure that random properties of a $j$th unitary, e.g. symmetry, do not lead to biased results. For 'Bristol' always $j=1000$ matrices are sampled due to the dispersed results. Subsequently the full set of primary data, $\left(\boldsymbol{\mathcal{V}},\boldsymbol{\tau},\boldsymbol{\theta}\right)_{m,j}$, is computed from each $\mathbf{H}_{m,j}$, where $\boldsymbol{\mathcal{V}}$, $\boldsymbol{\tau}$, and $\boldsymbol{\theta}$ denote the sets of two-photon visibilities, transmission intensities and phases sensed via coherent states, respectively. Under experimental conditions the primary data sets would be afflicted by statistic and systematic noise. We mimic this by perturbing the primary data sets with noise drawn from a normal distribution $\mathcal{N}(0,\,\sigma^2)$ of standard deviation $\sigma$ centred around zero. The perturbed primary data distributions are given as $(\boldsymbol{\tilde{\mathcal{V}}},\boldsymbol{\tilde{\tau}})_{m,j,\sigma}=(1+\mathcal{N}(0,\,\sigma^2))\left(\boldsymbol{\mathcal{V}},\boldsymbol{\tau}\right)_{m,j}$ and $20$ different values of $\sigma$ are sampled in $0.5\%$ steps from $\sigma=0.5\%$ to $\sigma=10\%$. Note that for the phase data ,$\boldsymbol{\tilde{\theta}}_{m,j,\sigma}=\boldsymbol{\theta}_{m,j}+\mathcal{N}(0,\,\sigma^2)$, absolute perturbances were chosen. Eventually the unitary descriptions are calculated via a Monte Carlo method drawing the data required for each reconstruction approach, $\mu$, randomly from $(\boldsymbol{\tilde{\mathcal{V}}},\boldsymbol{\tilde{\tau}},\boldsymbol{\tilde{\theta}})_{m,j,\sigma}$. An average unitary description, $\boldsymbol{\tilde{U}}_{m,j,\sigma,\mu}$, is obtained after $120$ iterations with $\sigma(\boldsymbol{\tilde{U}}_{m,j,\sigma,\mu})$ denoting its standard deviation. This way errors are estimated via an identical procedure, independent of whether an analytic error propagation method is available or not. Finally the fidelity (see eq. \ref{eq:fid}) for each of the $\approx 10^5$ reconstructed unitaries is computed.\\
A subset of the computed data is shown in Figure~\ref{fig:num_eval}\textbf{b)}, visualized as a two dimensional histogram for $m=4$ and $\mu=\text{'Brisbane'}$. Here the data points along one row, i.e for a given perturbance $\sigma$, are composed of $j=1000$ instead of $j=120$ fidelities, for visualization purposes only. The absolute frequencies for a given $\sigma$ can be associated with a probability distribution of a certain width, where the highest peak represents the most probable fidelity. For small perturbances $\sigma$ those distributions will be in general sharp but asymmetric, whereas for larger $\sigma$ dispersed and more symmetric distributions are found. To capture all but the most dispersed results we chose to fit them with a Weibull distribution centred around the most probable value, $\mathbf{\tilde{F}}_{m,\sigma,\mu}$. As an error measure, $\sigma_{\frac{1}{e}}(\mathbf{\tilde{F}}_{m,\sigma,\mu})$, the distances between the most probable fidelity and the two fidelities where the maximum probability decreased to $\frac{1}{e}$ are used.\\

\begin{figure}[h!]
\centering
\includegraphics[width=0.98\textwidth]{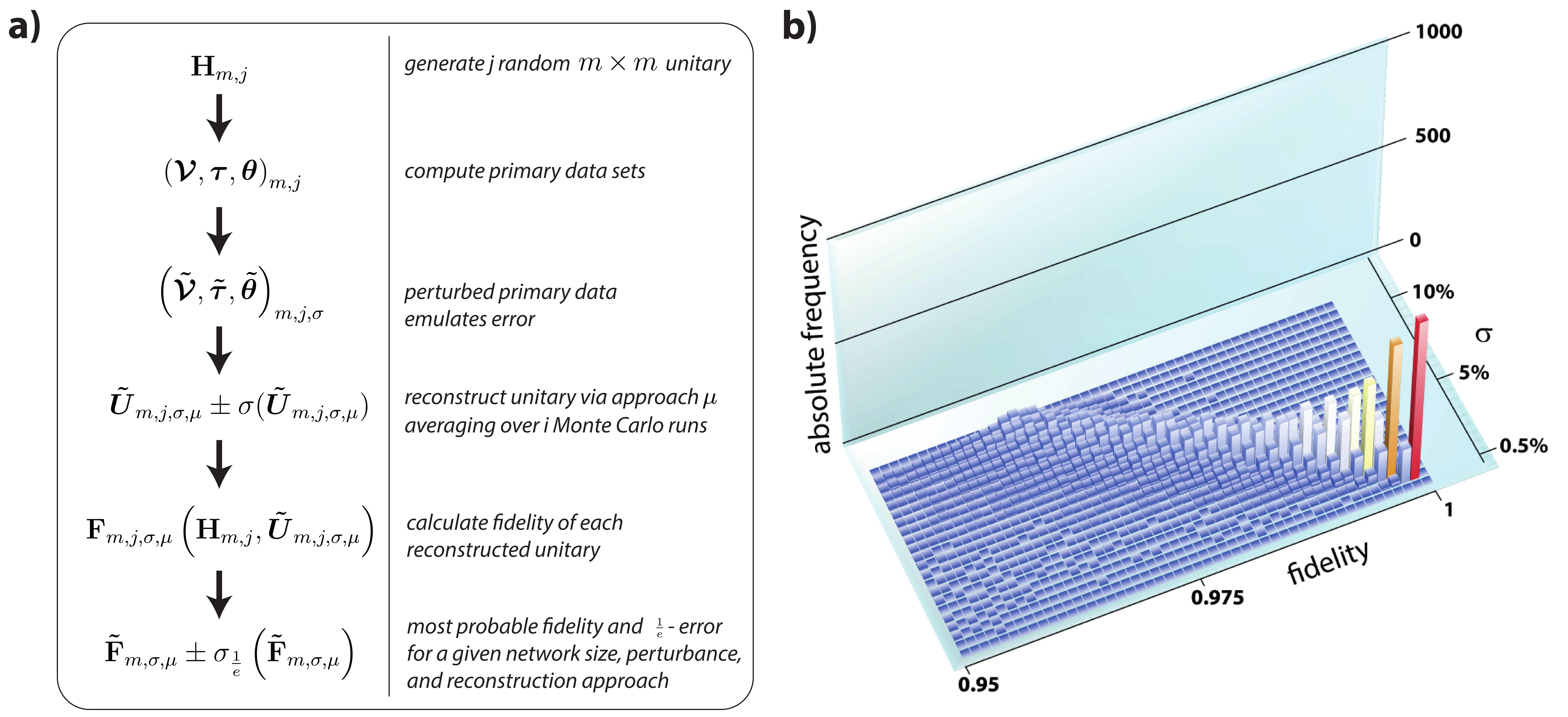}
\caption{\textbf{a) Flowchart} for the numerical method used to evaluate the different reconstruction approaches. \textbf{b) Frequency histogram} of fidelities obtained in the case that the reconstruction approach 'Brisbane' is applied to $4\times4$ networks. The fidelity axis is divided into $50$ bins ranging from $0.95$ to $1$, whereas the perturbance on the primary data, $\sigma$, ranges from $0.5\%$ to $10\%$ in $0.5\%$ steps. For illustration purposes the sampling size is increased from $120$ to $1000$ $4\times4$ Haar random unitaries for each $\sigma$.} 
\label{fig:num_eval}
\end{figure}

\textbf{Probability distributions}\\
\\
For the lossless case discussed above, Weibull distributions are used to extract the most probable fidelity, $\mathbf{\tilde{F}}_{m,\sigma,\mu}$, and the $\frac{1}{e}$ errors of the fidelity, $\sigma_{\frac{1}{e}}(\mathbf{\tilde{F}}_{m,\sigma,\mu})$. The Weibull distribution is given as
\begin{equation*}
f_{\text{Wb}}(x;k,\lambda)=\frac{k}{\lambda} \left(\frac{x}{\lambda}\right)^{k-1} e^{-(\frac{x}{\lambda})^k}\,,
\end{equation*}

with $\lambda$ and $k$ the scale and shape parameter, respectively.\\

\section*{Appendix B: The generation of primary data}\label{sec:II}

An $m$-mode linear optical scattering network can imprint new amplitude and phase information on an impinging light field (see figure~\ref{FIG:Sensing}). Whereas a large class of classical and quantum light fields can be used with the integrated optical networks considered here, their main application lies in the manipulation of coherent states or single photon Fock-states. Likewise, both states of light are suited as probe states to sense the transition-amplitudes or phases imprinted by the network. In the case that the light source used for characterization differs from the light source used in an experiment care has to be taken that the physical properties, especially the frequency and frequency bandwidth, are kept identical.\\
When injected into a single input port $k$ of a network, both coherent- and single-photon Fock-states allow to sense the transition amplitude $\tau_{j,k}$ of a specific matrix entry $U_{j,k}$ with $j$ denoting an output mode\footnote{The recorded intensities are proportional to $|\tau_{j,k}|^2$ with $\tau_{j,k}\geq 0$}. However, intensities measured in this way will be affected by loss, coupling and detection efficiencies. Rudimentary techniques to directly measure input loss~\cite{Rahimi-Keshari2013} work only in the case of zero output loss. Alternatively input and output loss can be traced out during the reconstruction process~\cite{Laing2012}. While this procedure is under ideal conditions loss-insensitive, the required algebraic transformations may even amplify error stemming from the primary data. Thus loss still presents a major problem when sensing transition-amplitudes and is best dealt with by careful calibration of e.g. detector efficiencies. This way a complete and reasonably accurate set of $m^2$ real entries of any $m\times m$ unitary can be sensed if mode-dependent propagation loss plays a secondary role.\\
To sense the phases of a linear optical network coherent states can be distributed among two input modes $k$ and $l$, $|\alpha_1\rangle_k|e^{i\varphi}\alpha_2\rangle_l$~\cite{Rahimi-Keshari2013}. It is sufficient to choose $l=1$ and subsequently measure the different input combinations $k=2\ldots m$. Modulating the phase $\varphi$ of this two-mode state at a frequency $\omega$ results in output intensities in all coupled output modes that are subjected to the modulation frequency $\omega$ albeit featuring a relative phase shift $\gamma_{j,k}$ between output modes. These relative phase shifts correspond to the phases $\gamma_{j,k}$ of the unitary matrix entry $U_{j,k}$ with all $\gamma_{j,k}=0$ for $j\vee k=1$ and an arbitrary sign for $\gamma_{2,2}$. Omitting the intensities and only recording this relative phase renders the measurement insensitive to mode-dependent input and output loss.\\
Experimentally the modulation $\omega$ can be realized through a piezo actuated mirror, a delay line or similar devices. Since coupling to integrated devices is predominantly implemented with fiber arrays the phase $\varphi$ and modulation with frequency $\omega$ will be affected by fluctuations caused by temperature or vibrations~\cite{Musha1982,hati2011vibration}. Care has to be taken that such noise is kept below a threshold which still allows to identify a $\omega$-periodicity in the output signal. In general the error can be largely minimized by utilizing a modulation frequency $\omega$ that is well separated from the frequency of the noise in the laboratory.\\
An alternative technique to sense the phase information utilizes the non-classical interference of two photons~\cite{Hong1987}. It can be shown~\cite{Peruzzo2011,Laing2012} that the extend of this quantum effect, the visibility of the resulting interference curve, is sensitive to the phases $\gamma_{j,k}$ of the interferometric network. Only the special case of a $2\times2$ device is phase-insensitive, owed to the unitary scattering submatrix. In general phase-sensitive probe states are also transition-amplitude sensitive and are therefore sufficient to generate all primary data needed to reconstruct the unitary description of a linear optical network. Remarkably, the visibilities obtained via two-photon interferences are insensitive to input and output loss. In contrast, sensing transition-amplitudes with two-mode coherent states generates the same problems as measuring transition amplitudes directly.\\
Experimentally, non-classical interference visibility measurements are ideally implemented using a pure, separable bi-photon state, where each photon is injected into its own interferometer mode. Subsequently the distinguishability of the two photons is scanned, e.g. by altering the relative temporal delay $\Delta\tau$ between the photons. Given that coupling to waveguides and propagation in waveguides is not lossless and that detection efficiency of e.g. avalanche photo diodes is limited, measurement times exceed those of coherent probe-states. Imperfections in the probe-state generation are the major contributes to systematic errors and affect the quality of the measured visibility. Using an involved modelling contributions from spectral mismatch and spectral correlations, background noise and drift in the coupling can be taken into account. Thus the accuracy of the extracted visibilities can be increased and errors minimized to $\approx 10^{-2}$. Complete sampling of all accessible visibilities, $N_{Vis}=\binom{m}{2}^2$ for a $m\times m$ network has several algorithmic and statistic advantages and can be reduced to $N_{Vis_{red}}=\binom{m}{2}$ measurement runs if a sufficient number of detectors is available. In many quantum optical experiments generating the data via the non-classical interference of two photons has the additional benefit that the apparatus required for characterization of the network is a subset of the whole experimental apparatus and the procedure works 'in-situ'. 

\begin{figure*}[ht]
\includegraphics[width=0.975\textwidth]{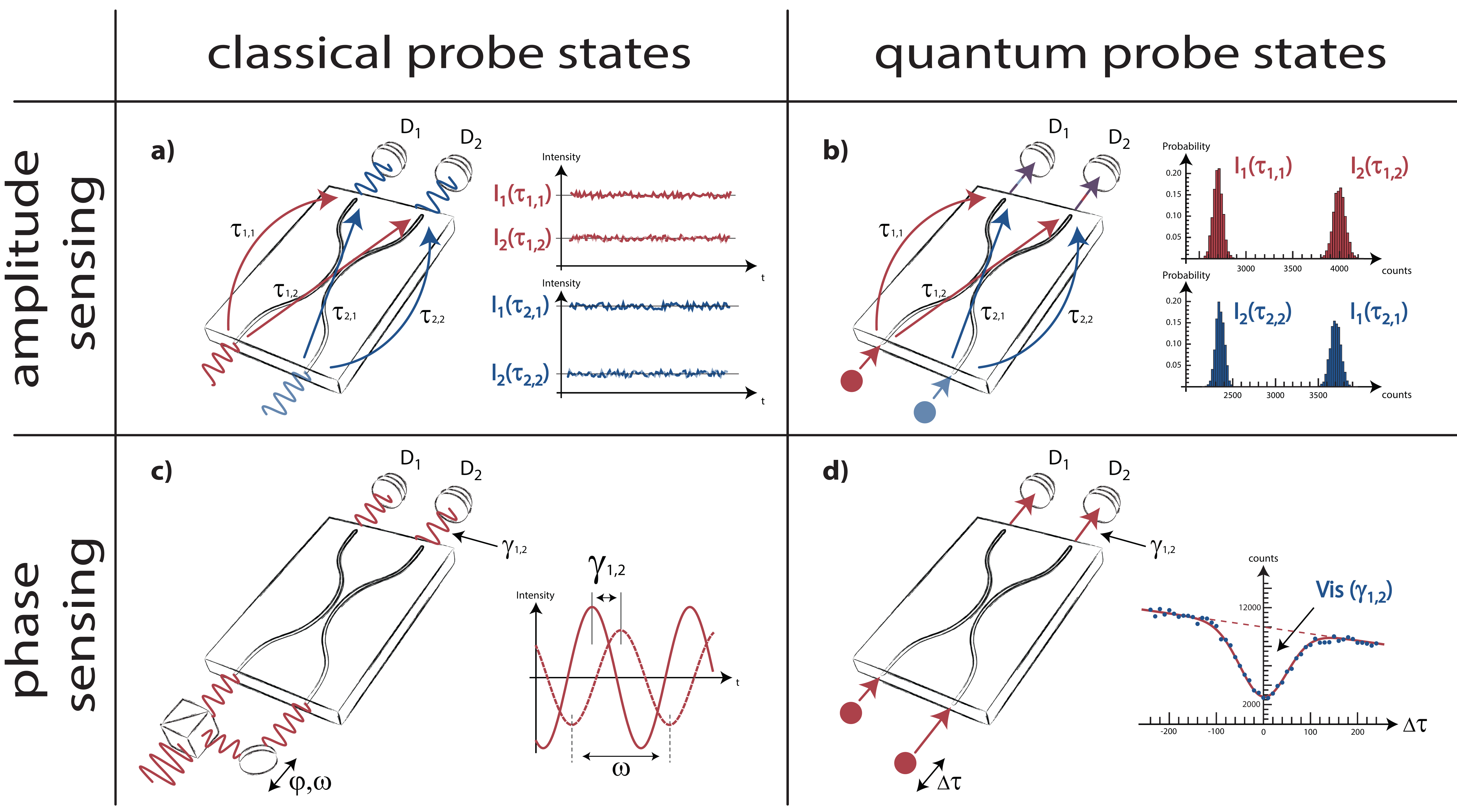}
\caption{\textbf{Classical and quantum probe states for transition amplitude and phase sensing.} Sensing the properties of linear optical interferometers is best understood by considering $2\times 2$ scattering submatrices of a larger interferometer. Here, both classical, \textbf{a)} and \textbf{c)}, and quantum, \textbf{b)} and \textbf{d)}, states of light can be used to sense information about the implemented amplitude and phase transformation. \textbf{a)} Coherent states and \textbf{b)} single-photon Fock-states injected into one mode of the network allow to measure the transmission and reflection intensity $I_1(\tau_{1,1})$ and $I_2(\tau_{1,2})$ (shown in red). Repeating the measurement via a second input port (shown in blue) allows to derive a loss-insensitive splitting parameter. Probing of non-global phases relies on interferometry which allows to extract such phases from intensity measurements. \textbf{c)} A coherent state is distributed among two input modes and the relative phase $\varphi$ is e.g. linearly modulated with frequency $\omega$. Here the phase $\gamma_{1,2}$ of the interferometer manifests as the relative phase of the recorded intensity pattern. \textbf{d)} In the case of $2\times 2$ scattering submatrices the Hong-Ou-Mandel Dip can be utilized to sense the phase $\gamma_{1,2}$ via the visibility $Vis({\gamma_{1,2}})$ of the two-photon interference curve, as these $2\times 2$ scattering submatrices are in general non-unitary. In contrast, monolithic $2\times 2$ blocks, i.e. a beam splitter, are unitary and hence the visibility is only affected by the splitting ratio.}
\label{FIG:Sensing}
\end{figure*}

\section*{APPENDIX C: Size of the primary data set and resource cost}\label{sec:cost}
Through technological progress the number of fully coupled modes supported by integrated circuits is growing and consequently, so is the size of the primary data sets required to characterize their unitary transformations. Thus a manageable size of these data sets is becoming an important criterion for evaluating different reconstruction approaches. Here, fully coupled interferometers represent the most general case. The lower bound of required data points is quantified by the number $n_{min}=2\binom{m}{2}$ of spherical coordinates that parametrize such a $m\times m$ network, with the spherical coordinates corresponding to the beam splitting ratios and phase shifts of Reck et. al~\cite{Reck1994}. Note that both, 'Bristol' and 'Brisbane' aim to reconstruct the $2m^2$ matrix entries directly, hence a set of $n_{min}$ data points is insufficient. The approaches 'Bristol' and 'Brisbane' utilize sufficient data sets that are close to this lower bound and consist of $m^2$ transition amplitudes and $(m-1)^2$ data points to recover the non-trivial phases. In the case of 'Bristol' additional $(m-1)^2-1$ two-photon visibilities are required to determine the sign of the phases. We chose the upper bound of primary data to be the over-complete set of all two-photon interference visibilities, $n_{\text{full}}=\binom{m}{2}^2$. This set of data can be efficiently recorded given today’s bright single photon sources~\cite{Jin2014} but can be in principle expanded to even higher order correlation functions. Likewise the $m^2$ transition amplitudes represent a non-redundant set of data to expand $n_{full}$. Since the transition amplitudes are loss afflicted they are not used in the global optimization routine of 'Vienna'. Only in the case that 'Vienna' is applied to black-box networks the $m^2$ transition amplitudes and $(m-1)^2$ relative phases are needed to extract the starting parameters for the global optimization. Furthermore this global optimization allows for an adaptive strategy; best reconstruction accuracy is achieved using the full over-complete set of data. Alternatively, a reduced set of data, the set of all possible two-photon interference visibilities that can be generated when one photon is always inserted into input port one and the second photon into input port $2,\dots,m$, can be used. This results in a reduced set of $n_{Vienna_{min}}=(m-1)\binom{m}{2}$ visibilities which always suffices to reconstruct $m\times m$ networks with $m\geq 3$ modes. In the following we will refer to the reconstruction approach 'Vienna' utilizing a reduced set of primary data as 'Vienna reduced'.\\
A second number, the required measurement runs to generate the primary data set, 
can be regarded as the experimentally more relevant parameter. We list this parameter for the reconstruction approaches compared here in the limit that every output mode is coupled to an individual detector in table~\ref{tab:cost2}. Now, all output events for a given input combination can be recorded in parallel, thus the number of required measurements  corresponds to the required input combinations that need to be consecutively aligned in a laboratory. For instance, the $m^2$ transition amplitude data can be acquired in $m$ measurement runs and the $\binom{m}{2}^2$ two-photon interference visibilities in $\binom{m}{2}$ measurement runs.

\begin{table}[ht]
\begin{tabular}{|>{\centering\arraybackslash}m{4cm}|c|c|c|c|}
\hline  & 'Brisbane' & 'Bristol' & 'Vienna' & 'Vienna$_{\text{black-box}}$'\\
\hline \rule[-2ex]{0pt}{5.5ex} minimal primary data used & $m^2+(m-1)^2$ & $m^2+(m-1)^2+(m-1)^2-1$ & $(m-1)\binom{m}{2}$ & $m^2+(m-1)^2+(m-1)\binom{m}{2}$ \\
\hline \rule[-2ex]{0pt}{5.5ex} required number of measurement runs with sufficient detectors  & $2m-1$  & $3m-3$ & $m-1$ & $3m-2$\\
\hline
\hline \rule[-2ex]{0pt}{5.5ex} maximal available primary data & $m^2+(m-1)^2$ & $m^2+\binom{m}{2}^2$ & $\binom{m}{2}^2$ & $m^2+(m-1)^2+\binom{m}{2}^2$ \\
\hline \rule[-2ex]{0pt}{5.5ex} required number of measurement runs with sufficient detectors & $2m-1$ &$m+\binom{m}{2}$  & $\binom{m}{2}$ & $(2m-1)+\binom{m}{2}$ \\
\hline
\end{tabular}
\caption{\textbf{Size of the primary data set and minimal number of measurements for different reconstruction approaches.} The three compared unitary reconstruction approaches differ significantly in the minimal and maximal set of primary data available for the reconstruction algorithms. For 'Brisbane' the minimal and maximal set of data are identical. Whereas a larger primary data set is more costly to generate experimentally this expense is justified if the data can be used to increase the accuracy of the reconstructed description. The required number of measurement runs is given in the limit that each output mode is covered by its own detector and can be regarded as the experimentally more relevant quantity. Here 'Vienna' is referring to a reconstruction of a structure with known layout and starting values, while 'Vienna$_{\text{black-box}}$' refers to a black-box network. The minimal primary data set required for the approach 'Bristol' presents a special case. Here the amount of data is constituted by the $m^2$ transition amplitudes and the $(m-1)^2$ and $(m-1)^2-1$ two-photon visibilities to recover the absolute values and signs of the phases, respectively. In the case of sufficient detectors the transition amplitudes can be sensed in $m$ measurement runs and the two-photon visibilities to recover the absolute value of the phases in $m-1$ measurement runs. To fix the sign of the phases additional $m-2$ measurement runs are necessary.}
\label{tab:cost2}
\end{table}

\section*{APPENDIX D: The influence of loss}\label{sec:loss}
Loss can be a major factor in experiments using integrated optical circuits. In the case of direct laser-written networks propagation loss of $-0.3\frac{\text{dB}}{cm}$ and coupling loss of $-2\text{dB}$ are typical values~\cite{Lebugle2015,Bentivegna2015}. If these losses are mode-independent, however, they just represent a global loss term that commutes with the optical transformation of a LON. In contrast, mode-dependent losses cause deviations in a targeted optical transformation. Characterizing mode-dependent losses thus becomes important when reconstructing the optical transformation of a LON. Here, we distinguish between the case of mode-dependent loss at the input and output ports and the case of mode-dependent propagation loss. In the first case, loss can always be separated from the transformation of a LON and be described by loss matrices containing virtual beam splitters. For a $m \times m$ LON this translates to $m$ additional parameters modelling input loss and $m$ additional parameters modelling output loss, the $\alpha_{i}$ and $\alpha_{o}$ of figure~\ref{FIG:4times4loss}, respectively. The unitary matrix of the LON expanded by the $2m$ loss modes now reads as

\begin{equation}\label{eq:Ulossy}
\boldsymbol{U}_{3m \times 3m}=
\boldsymbol{L}(\boldsymbol{\alpha_o})\times
 \begin{pmatrix} 
\boldsymbol{U}_{m \times m} & 0 \\
0 & \boldsymbol{I}_{2m \times 2m} \\
 \end{pmatrix}
\times \boldsymbol{L}(\boldsymbol{\alpha_i}),
\end{equation}

with $\boldsymbol{L}$ denoting the $3m \times 3m$ loss matrices. Still, only the original $m$ input and output modes are experimentally accessible. Note that the data set for relative phases sensed via coherent states, $\boldsymbol{\theta}$, and the the data set composed of two-photon interference visibilities, $\boldsymbol{\mathcal{V}}$, are insensitive to these losses and hence directly reveal information about $\boldsymbol{U}_{m \times m}$. Whereas $\boldsymbol{\theta}$ just contains information on the non-trivial phases of the interferometer but not on the transition amplitudes, the set of two-photon interference visibilities contains information on both. This is a unique feature of the latter data set and owed to the quantum nature of the interference.

\begin{figure*}[ht]
\includegraphics[width=0.975\textwidth]{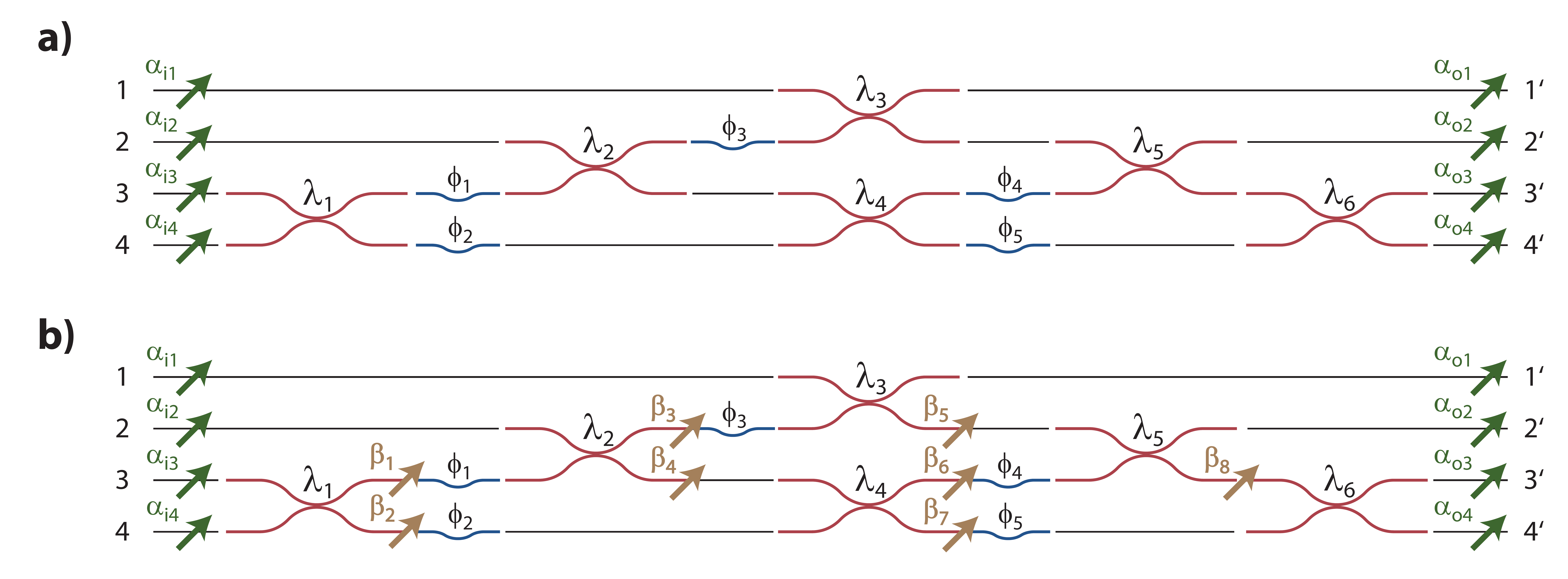}
\caption{\textbf{Layout of $4\times 4$ interferometers} following the Reck et al. scheme. Tuning the splitting ratios, $\lambda_i$, of the $\binom{m}{2}$ beam splitters and the phase shifts, $\phi_i$, of the $\binom{m}{2}-1$ phase shifters any $4\times4$ unitary can be realized. Phases at the input and output, i.e. global phases, are omitted. \textbf{a)} Input and output loss can be parametrized by eight additional beam splitters, $\alpha_{i_i}$ and $\alpha_{o_i}$, coupling to the loss modes $5$ to $12$, here indicated by the green arrows. \textbf{b} Mode-dependent propagation loss can be modelled with additional $2 \binom{m}{2}-m$ beam splitters, $\beta_{i}$, coupling to the loss modes $13$ to $20$, here indicated by the light brown arrows.}
\label{FIG:4times4loss}
\end{figure*}

In comparison, a direct measurement of the transition amplitudes $\boldsymbol{\tau}$ is always afflicted by mode-dependent input and output loss. As a consequence, the latter set of data only reveals information on a $m \times m$ submatrix of $\boldsymbol{U}_{3m \times 3m}$. In our notation this submatrix corresponds to the upper left $m \times m$ submatrix of $\boldsymbol{U}_{3m \times 3m}$, that is spanned by the $m$ accessible input and output ports. In general, the size of the transition amplitude data set $\boldsymbol{\tau}$ is insufficient to reconstruct all the $2m$ loss terms in addition to the $m^2$ real entries directly. Therefore, the strategy used in 'Brisbane' only works if $m$ loss parameters, either all $\alpha_i$ or all $\alpha_o$, can be neglected (see also body text). Alternatively, the transition amplitude data can be subjected to a reconstruction algorithm in which the loss terms drop out, as done in 'Bristol', however this necessitates unitary constraints. Although the strategy of 'Bristol' seems to be of advantage, figure~\ref{results:comparison} gives a indication that the algorithm reacts fragile to measurement error on the primary data. In comparison 'Brisbane' achieves higher reconstruction fidelities. This may change in the presence of mode-dependent input and output loss. It is an open question where the threshold of measurement error on the sensed data opposed to the level of of input and output loss lies, that would favour one over the other algorithm. Due to the processing of just two-photon interference visibilities, 'Vienna' is insensitive to input and output losses. Hence, $\boldsymbol{U}_{m \times m}$ is directly reconstructed and the $\alpha_i$ and $\alpha_o$ can be obtained in a separate step by solving a system of linear equations using loss afflicted data, e.g. the set of transition amplitudes $\boldsymbol{\tau}$. In the case of black-box networks, however, some dependence is carried over if initial starting parameters, $\boldsymbol{p}$, are obtained via, e.g. 'Brisbane'.\\
In contrast to input and output loss, mode-dependent propagation loss cannot be separated from the unitary description of a LON, as it does not commute with the optical elements that constitute the network's fundamental transformation. Instead, the unitary description needs to be expanded by additional in-circuit loss modes, subsequently labelled $l$. This is illustrated in figure~\ref{FIG:4times4loss}b) for the case of a $4\times4$ circuit, where the $\beta_i$ denote the parameters of the loss modelling beam splitters. For general $m\times m$ LONs which follow the Reck et al. layout this translates to $l=2\binom{m}{2}-m$ additional modes. Now the unitary can be written as

\begin{equation}\label{eq:Ufulllossy}
\boldsymbol{U}_{r \times r}=
\boldsymbol{L'}(\boldsymbol{\alpha_o})\times
 \begin{pmatrix} 
\boldsymbol{U}_{(m+l) \times (m+l)} & 0 \\
0 & \boldsymbol{I}_{2m \times 2m} \\
 \end{pmatrix}
\times \boldsymbol{L'}(\boldsymbol{\alpha_i}),
\end{equation}

with $m$ the number of accessible network-modes, $r=3m+l$ the total number of modes and $\boldsymbol{L'}$ denoting the $r \times r$ input and output loss matrices. All data sets sensed for reconstruction purposes, $\boldsymbol{\tau}\,,\boldsymbol{\theta}\,,\boldsymbol{\mathcal{V}}$ reveal information about an in general non-unitary submatrix of $\boldsymbol{U}_{r \times r}$. In our notation, this submatrix corresponds to the upper left $m \times m$ submatrix of $\boldsymbol{U}_{r \times r}$. Now, the above mentioned strategies to model loss fail. The loss terms cannot be assessed directly and error is introduced by applying unitary constrains to the non-unitary $m \times m$ submatrix.\\ 
We numerically evaluate the exemplary case of $m=4$ LONs to investigate how severe the reconstruction performance of the various approaches is offset by mode-dependent in-circuit loss. The general layout of these networks is shown in figure~\ref{FIG:4times4loss}\textbf{b)}. Here, the input and output losses are kept zero to ensure that they do not influence the results. Thus the networks are just expanded by $l=8$ in-circuit loss modes.
The approaches 'Brisbane' and 'Bristol' are constructed to obtain $4\times4$ descriptions which prevents the use of the fidelity measure defined in equation~\ref{eq:fid}. Hence we use an alternative measure which is experimentally motivated and constructed as the mean deviation between a point of primary data and its prediction obtained via one of the reconstructed descriptions. $\mathcal{Q}_{\text{t}}$ quantifies the mean deviation for the normalized transition-amplitude data and $\mathcal{Q}_{\text{vis}}$ the mean deviation for the two-photon interference visibilities (see definition below). In the limit of $\mathcal{Q}_{\text{t}}=\mathcal{Q}_{\text{vis}}=0$, perfect reconstruction is achieved, a result only expected if in-circuit loss is either zero or if it can be fully recovered by an approach. One option to achieve the latter is a reconstruction of the complete $(4+8)\times(4+8)$ unitary description, which we investigate for the approach 'Vienna'. The required starting parameters for the $\lambda_i$ and $\phi_i$ used to initialize to optimization routine are extracted via the approach 'Brisbane'. All $\beta_i$ are initialized at zero. The data presented in figure~\ref{FIG:LossyResults} is sampled for different levels of loss by drawing the transmittances of the loss beam splitters, $\beta_1, \ldots, \beta_8$, randomly from a uniform distribution $[\cos(\epsilon),1]$. In the worst case of $\epsilon=0.1$, the maximum loss per beam splitter thus is $\sin^2(0.1)\approx1\%$. To sample the unitary space representative, $j=500$ different $12\times12$ starting matrices are generated for each loss-interval.
The general perturbance on the primary data was set to $\sigma=1\%$. All reconstructed descriptions are computed in the same way as in section~\ref{sec:results} via a Monte Carlo method with a sampling size of $i=100$ (see also Appendix A). The resulting frequency histograms for $\mathcal{Q}_{\text{t}}$, $\mathcal{Q}_{\text{vis}}$ and the fidelity histograms in the case of 'Vienna' were fitted with Burr type XII distributions~\cite{Burr1942}, as these show good overlap with the numerical data. The data points and error bars contained within figure~\ref{FIG:LossyResults} are given as the most probable value of the distributions and the distance between the most probable value and those values to the left and right where the maximum probability decreases to $\frac{1}{e}$, respectively.\\
Already for the small levels of mode-dependent propagation loss considered here, all three approaches struggle to reconstruct precise unitary descriptions. This result is to be expected in the case of 'Brisbane' and 'Bristol' as the sets of data used for the reconstruction procedures originate from a non-unitary submatrix. Hence only closest unitary descriptions are reconstructed, which in turn cannot fully explain the original data sets.\\
The results obtained in the case of 'Vienna' need to be interpreted in a different way. Whereas the global optimization of just two-photon interference visibilities converges, as can be seen by the values for $\mathcal{Q}_{\text{vis}}$ in figure~\ref{FIG:LossyResults}\textbf{b)}, the values obtained for $\mathcal{Q}_{\text{t}}$ show a scaling similar to the other two approaches. This indicates that the original optical element parameters for the beam splitters, phase shifters and loss beam splitters cannot be recovered with high accuracy. As a result the fidelities of the reconstructed $12\times 12$ unitary matrices decrease rapidly with growing $\epsilon$. It is an open question whether a larger set of primary data including relative phase shifts $\boldsymbol{\tilde{\theta}}$ would yield improved reconstruction results. In light of the above results mode-dependent propagation loss still presents a fundamental problem when reconstructing unitary descriptions of LONs.\\

\begin{figure*}[ht]
\centering
\includegraphics[width=0.98\textwidth]{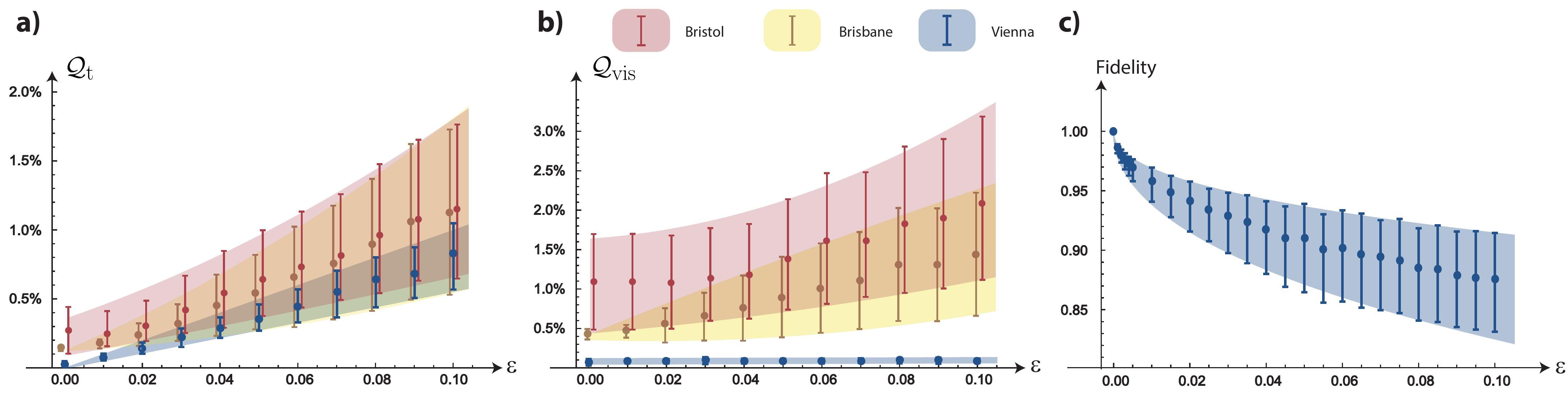}
\caption{\textbf{The influence of mode-dependent propagation loss on the reconstruction performance} for $4\times 4$ LONs and the different approaches 'Brisbane', 'Bristol', and 'Vienna'. The data for 'Brisbane' and 'Bristol' is offset from the data in the case of 'Vienna' for visualization purposes only. Here the general perturbance on the primary data, $\sigma$, was chosen to be $1\%$ and input and output loss to be zero. The transmittances, $\beta_1, \ldots, \beta_8$, of the eight beam splitters modelling the mode-dependent loss were drawn uniformly from the interval $[\cos(\epsilon),1]$. Several intervals were sampled in discrete steps ranging from zero loss to $\sin^2(\epsilon)=1\%$ loss and $500$ random matrices were reconstructed for each interval. The histograms were fitted with Burr type XII distributions and the error bars are given as the distance between the most probable value and those values to the left and right where the maximum probability decreases to $\frac{1}{e}$. \textbf{a)} $\mathcal{Q}_{\text{t}}$ quantifies the mean deviation of the normalized transition amplitudes between the initial values and the ones obtained from the reconstructed descriptions. \textbf{b)} Analogously $\mathcal{Q}_{\text{vis}}$ quantifies the mean deviation of the two-photon interference visibilities. Only data from the experimentally accessible $4\times 4$ submatrices is considered. \textbf{c)} Reconstruction fidelities for the $\boldsymbol{U}_{(4+8) \times (4+8)}$ unitary matrices reconstructed via 'Vienna'.}
\label{FIG:LossyResults}
\end{figure*}

 \textbf{Quality measure for the case of mode-dependent propagation loss}\\
\\
The quality measure $\mathcal{Q}_{\text{vis}}$ introduced in above is constructed as the difference between the full set of sensed two-photon interference visibilities and the set of predicted visibilities, obtained via a reconstructed unitary matrix. It is normalized by the maximum amount of two-photon interference visibilities, $\binom{m}{2}^2$, that can be obtained.
\begin{equation}
\mathcal{Q}_{\text{vis}}=\frac{1}{\binom{m}{2}^2}\sum_{i=1}^{\binom{m}{2}^2}\left|\text{Vis}_{i}^{\text{true}}-\text{Vis}_{i}^{\text{predicted}}\right|
\end{equation}
Similarly, the measure $\mathcal{Q}_{\text{t}}$ quantifies the difference between the set of measured transition-amplitudes and the predicted ones via a reconstructed unitary matrix. It is normalized by the maximum amount of transition-amplitude data, $m^2$, that can be obtained.
\begin{equation}
\mathcal{Q}_{\text{t}}=\frac{1}{m^2}\sum_{j=1}^{m}\sum_{k=1}^{m}\left|\tau_{j,k}^{*,\text{true}}-\tau_{j,k}^{*,\text{predicted}}\right|
\end{equation}
In the case of $\mathcal{Q}_{\text{t}}$, we normalize the transition-amplitude data for each of the $m$ measured inputs over all outputs to $m$ unit vectors, to allow for comparison between the loss sensitive and insensitive approaches. This normalized transition amplitude data is labelled $\tau^*$ and defined as
\begin{equation}
\left|\tau^*_{j,k}\right|=\frac{\left|\tau_{j,k}\right|}{\sum_{k=1}^{m}\left|\tau_{j,k}\right|}\,.
\end{equation}

\end{widetext}
\clearpage

\newpage


\begin{thebibliography}{10}
\expandafter\ifx\csname url\endcsname\relax
  \def\url#1{\texttt{#1}}\fi
\expandafter\ifx\csname urlprefix\endcsname\relax\def\urlprefix{URL }\fi
\providecommand{\bibinfo}[2]{#2}
\providecommand{\eprint}[2][]{\url{#2}}

\bibitem{Knill2001}
\bibinfo{author}{Knill, E.}, \bibinfo{author}{Laflamme, R.} \&
  \bibinfo{author}{Milburn, G.~J.}
\newblock \bibinfo{title}{A scheme for efficient quantum computation with
  linear optics}.
\newblock \emph{\bibinfo{journal}{nature}} \textbf{\bibinfo{volume}{409}},
  \bibinfo{pages}{46--52} (\bibinfo{year}{2001}).

\bibitem{Politi2009}
\bibinfo{author}{Politi, A.}, \bibinfo{author}{Matthews, J.~C.},
  \bibinfo{author}{Thompson, M.~G.} \& \bibinfo{author}{O'Brien, J.~L.}
\newblock \bibinfo{title}{Integrated quantum photonics}.
\newblock \emph{\bibinfo{journal}{IEEE Journal of Selected Topics in Quantum
  Electronics}} \textbf{\bibinfo{volume}{15}}, \bibinfo{pages}{1673--1684}
  (\bibinfo{year}{2009}).

\bibitem{Minkov2016}
\bibinfo{author}{Minkov, M.} \& \bibinfo{author}{Savona, V.}
\newblock \bibinfo{title}{A compact, integrated silicon device for the
  generation of spectrally filtered, pair-correlated photons}.
\newblock \emph{\bibinfo{journal}{Journal of Optics}}
  \textbf{\bibinfo{volume}{18}}, \bibinfo{pages}{054012}
  (\bibinfo{year}{2016}).

\bibitem{Zhang2016}
\bibinfo{author}{Zhang, X.}, \bibinfo{author}{Zhang, Y.},
  \bibinfo{author}{Xiong, C.} \& \bibinfo{author}{Eggleton, B.~J.}
\newblock \bibinfo{title}{Correlated photon pair generation in low-loss
  double-stripe silicon nitride waveguides}.
\newblock \emph{\bibinfo{journal}{Journal of Optics}}
  \textbf{\bibinfo{volume}{18}}, \bibinfo{pages}{074016}
  (\bibinfo{year}{2016}).

\bibitem{Silverstone2015}
\bibinfo{author}{Silverstone, J.} \emph{et~al.}
\newblock \bibinfo{title}{Qubit entanglement between ring-resonator photon-pair
  sources on a silicon chip}.
\newblock \emph{\bibinfo{journal}{Nature Communications}}
  \textbf{\bibinfo{volume}{6}} (\bibinfo{year}{2015}).

\bibitem{Harris2014}
\bibinfo{author}{Harris, N.~C.} \emph{et~al.}
\newblock \bibinfo{title}{Integrated source of spectrally filtered correlated
  photons for large-scale quantum photonic systems}.
\newblock \emph{\bibinfo{journal}{Physical Review X}}
  \textbf{\bibinfo{volume}{4}}, \bibinfo{pages}{041047} (\bibinfo{year}{2014}).

\bibitem{Konoike2016}
\bibinfo{author}{Konoike, R.} \emph{et~al.}
\newblock \bibinfo{title}{On-demand transfer of trapped photons on a chip}.
\newblock \emph{\bibinfo{journal}{Science Advances}}
  \textbf{\bibinfo{volume}{2}}, \bibinfo{pages}{e1501690}
  (\bibinfo{year}{2016}).

\bibitem{Corrielli2016}
\bibinfo{author}{Corrielli, G.}, \bibinfo{author}{Seri, A.},
  \bibinfo{author}{Mazzera, M.}, \bibinfo{author}{Osellame, R.} \&
  \bibinfo{author}{de~Riedmatten, H.}
\newblock \bibinfo{title}{Integrated optical memory based on laser-written
  waveguides}.
\newblock \emph{\bibinfo{journal}{Physical Review Applied}}
  \textbf{\bibinfo{volume}{5}}, \bibinfo{pages}{054013} (\bibinfo{year}{2016}).

\bibitem{Najafi2015}
\bibinfo{author}{Najafi, F.} \emph{et~al.}
\newblock \bibinfo{title}{On-chip detection of non-classical light by scalable
  integration of single-photon detectors}.
\newblock \emph{\bibinfo{journal}{Nature communications}}
  \textbf{\bibinfo{volume}{6}} (\bibinfo{year}{2015}).

\bibitem{Kahl2015}
\bibinfo{author}{Kahl, O.} \emph{et~al.}
\newblock \bibinfo{title}{Waveguide integrated superconducting single-photon
  detectors with high internal quantum efficiency at telecom wavelengths}.
\newblock \emph{\bibinfo{journal}{Scientific reports}}
  \textbf{\bibinfo{volume}{5}} (\bibinfo{year}{2015}).

\bibitem{Akhlaghi2015}
\bibinfo{author}{Akhlaghi, M.~K.}, \bibinfo{author}{Schelew, E.} \&
  \bibinfo{author}{Young, J.~F.}
\newblock \bibinfo{title}{Waveguide integrated superconducting single-photon
  detectors implemented as near-perfect absorbers of coherent radiation}.
\newblock \emph{\bibinfo{journal}{Nature communications}}
  \textbf{\bibinfo{volume}{6}} (\bibinfo{year}{2015}).

\bibitem{Pernice2012}
\bibinfo{author}{Pernice, W.~H.} \emph{et~al.}
\newblock \bibinfo{title}{High-speed and high-efficiency travelling wave
  single-photon detectors embedded in nanophotonic circuits}.
\newblock \emph{\bibinfo{journal}{Nature communications}}
  \textbf{\bibinfo{volume}{3}}, \bibinfo{pages}{1325} (\bibinfo{year}{2012}).

\bibitem{Politi2008}
\bibinfo{author}{Politi, A.}, \bibinfo{author}{Cryan, M.~J.},
  \bibinfo{author}{Rarity, J.~G.}, \bibinfo{author}{Yu, S.} \&
  \bibinfo{author}{O'Brien, J.~L.}
\newblock \bibinfo{title}{Silica-on-silicon waveguide quantum circuits}.
\newblock \emph{\bibinfo{journal}{Science}} \textbf{\bibinfo{volume}{320}},
  \bibinfo{pages}{646--649} (\bibinfo{year}{2008}).

\bibitem{Marshall2009}
\bibinfo{author}{Marshall, G.~D.} \emph{et~al.}
\newblock \bibinfo{title}{Laser written waveguide photonic quantum circuits}.
\newblock \emph{\bibinfo{journal}{Optics Express}}
  \textbf{\bibinfo{volume}{17}}, \bibinfo{pages}{12546--12554}
  (\bibinfo{year}{2009}).

\bibitem{Crespi2011}
\bibinfo{author}{Crespi, A.} \emph{et~al.}
\newblock \bibinfo{title}{Integrated photonic quantum gates for polarization
  qubits}.
\newblock \emph{\bibinfo{journal}{Nature communications}}
  \textbf{\bibinfo{volume}{2}}, \bibinfo{pages}{566} (\bibinfo{year}{2011}).

\bibitem{Heilmann2014}
\bibinfo{author}{Heilmann, R.}, \bibinfo{author}{Gr{\"a}fe, M.},
  \bibinfo{author}{Nolte, S.} \& \bibinfo{author}{Szameit, A.}
\newblock \bibinfo{title}{Arbitrary photonic wave plate operations on chip:
  Realizing hadamard, pauli-x, and rotation gates for polarisation qubits}.
\newblock \emph{\bibinfo{journal}{Scientific reports}}
  \textbf{\bibinfo{volume}{4}} (\bibinfo{year}{2014}).

\bibitem{Corrielli2014}
\bibinfo{author}{Corrielli, G.} \emph{et~al.}
\newblock \bibinfo{title}{Rotated waveplates in integrated waveguide optics}.
\newblock \emph{\bibinfo{journal}{Nature communications}}
  \textbf{\bibinfo{volume}{5}} (\bibinfo{year}{2014}).

\bibitem{Thompson2011}
\bibinfo{author}{Thompson, M.~G.}, \bibinfo{author}{Politi, A.},
  \bibinfo{author}{Matthews, J.~C.} \& \bibinfo{author}{O'Brien, J.~L.}
\newblock \bibinfo{title}{Integrated waveguide circuits for optical quantum
  computing}.
\newblock \emph{\bibinfo{journal}{IET circuits, devices \& systems}}
  \textbf{\bibinfo{volume}{5}}, \bibinfo{pages}{94--102}
  (\bibinfo{year}{2011}).

\bibitem{Shadbolt2012}
\bibinfo{author}{Shadbolt, P.~J.} \emph{et~al.}
\newblock \bibinfo{title}{Generating, manipulating and measuring entanglement
  and mixture with a reconfigurable photonic circuit}.
\newblock \emph{\bibinfo{journal}{Nature Photonics}}
  \textbf{\bibinfo{volume}{6}}, \bibinfo{pages}{45--49} (\bibinfo{year}{2012}).

\bibitem{Metcalf2014}
\bibinfo{author}{Metcalf, B.~J.} \emph{et~al.}
\newblock \bibinfo{title}{Quantum teleportation on a photonic chip}.
\newblock \emph{\bibinfo{journal}{Nature Photonics}}
  \textbf{\bibinfo{volume}{8}}, \bibinfo{pages}{770--774}
  (\bibinfo{year}{2014}).

\bibitem{Carolan2015}
\bibinfo{author}{Carolan, J.} \emph{et~al.}
\newblock \bibinfo{title}{Universal linear optics}.
\newblock \emph{\bibinfo{journal}{Science}}  (\bibinfo{year}{2015}).

\bibitem{Lebugle2015}
\bibinfo{author}{Lebugle, M.} \emph{et~al.}
\newblock \bibinfo{title}{Bloch oscillations of einstein-podolsky-rosen
  states}.
\newblock \emph{\bibinfo{journal}{arXiv preprint arXiv:1501.01764}}
  (\bibinfo{year}{2015}).

\bibitem{Martinis2015}
\bibinfo{author}{Martinis, J.~M.}
\newblock \bibinfo{title}{Qubit metrology for building a fault-tolerant quantum
  computer}.
\newblock \emph{\bibinfo{journal}{arXiv preprint arXiv:1510.01406}}
  (\bibinfo{year}{2015}).

\bibitem{Childs2001}
\bibinfo{author}{Childs, A.~M.}, \bibinfo{author}{Chuang, I.~L.} \&
  \bibinfo{author}{Leung, D.~W.}
\newblock \bibinfo{title}{Realization of quantum process tomography in nmr}.
\newblock \emph{\bibinfo{journal}{Physical Review A}}
  \textbf{\bibinfo{volume}{64}}, \bibinfo{pages}{012314}
  (\bibinfo{year}{2001}).

\bibitem{Mitchell2003}
\bibinfo{author}{Mitchell, M.}, \bibinfo{author}{Ellenor, C.},
  \bibinfo{author}{Schneider, S.} \& \bibinfo{author}{Steinberg, A.}
\newblock \bibinfo{title}{Diagnosis, prescription, and prognosis of a
  bell-state filter by quantum process tomography}.
\newblock \emph{\bibinfo{journal}{Physical review letters}}
  \textbf{\bibinfo{volume}{91}}, \bibinfo{pages}{120402}
  (\bibinfo{year}{2003}).

\bibitem{OBrien2004}
\bibinfo{author}{O'Brien, J.~L.} \emph{et~al.}
\newblock \bibinfo{title}{Quantum process tomography of a controlled-not gate}.
\newblock \emph{\bibinfo{journal}{Physical review letters}}
  \textbf{\bibinfo{volume}{93}}, \bibinfo{pages}{080502}
  (\bibinfo{year}{2004}).

\bibitem{Peruzzo2011a}
\bibinfo{author}{Peruzzo, A.}, \bibinfo{author}{Laing, A.},
  \bibinfo{author}{Politi, A.}, \bibinfo{author}{Rudolph, T.} \&
  \bibinfo{author}{O'Brien, J.~L.}
\newblock \bibinfo{title}{Multimode quantum interference of photons in
  multiport integrated devices}.
\newblock \emph{\bibinfo{journal}{Nature communications}}
  \textbf{\bibinfo{volume}{2}}, \bibinfo{pages}{224} (\bibinfo{year}{2011}).

\bibitem{Laing2012}
\bibinfo{author}{Laing, A.} \& \bibinfo{author}{O'Brien, J.~L.}
\newblock \bibinfo{title}{Super-stable tomography of any linear optical
  device}.
\newblock \emph{\bibinfo{journal}{arXiv preprint arXiv:1208.2868}}
  (\bibinfo{year}{2012}).

\bibitem{Lobino2008}
\bibinfo{author}{Lobino, M.} \emph{et~al.}
\newblock \bibinfo{title}{Complete characterization of quantum-optical
  processes}.
\newblock \emph{\bibinfo{journal}{Science}} \textbf{\bibinfo{volume}{322}},
  \bibinfo{pages}{563--566} (\bibinfo{year}{2008}).

\bibitem{Rahimi-Keshari2011}
\bibinfo{author}{Rahimi-Keshari, S.} \emph{et~al.}
\newblock \bibinfo{title}{Quantum process tomography with coherent states}.
\newblock \emph{\bibinfo{journal}{New Journal of Physics}}
  \textbf{\bibinfo{volume}{13}}, \bibinfo{pages}{013006}
  (\bibinfo{year}{2011}).

\bibitem{Rahimi-Keshari2013}
\bibinfo{author}{Rahimi-Keshari, S.} \emph{et~al.}
\newblock \bibinfo{title}{Direct characterization of linear-optical networks}.
\newblock \emph{\bibinfo{journal}{Optics express}}
  \textbf{\bibinfo{volume}{21}}, \bibinfo{pages}{13450--13458}
  (\bibinfo{year}{2013}).

\bibitem{Fedorov2015}
\bibinfo{author}{Fedorov, I.~A.}, \bibinfo{author}{Fedorov, A.~K.},
  \bibinfo{author}{Kurochkin, Y.~V.} \& \bibinfo{author}{Lvovsky, A.}
\newblock \bibinfo{title}{Tomography of a multimode quantum black box}.
\newblock \emph{\bibinfo{journal}{New Journal of Physics}}
  \textbf{\bibinfo{volume}{17}}, \bibinfo{pages}{043063}
  (\bibinfo{year}{2015}).

\bibitem{aaronson2011computational}
\bibinfo{author}{Aaronson, S.} \& \bibinfo{author}{Arkhipov, A.}
\newblock \bibinfo{title}{The computational complexity of linear optics}.
\newblock In \emph{\bibinfo{booktitle}{Proceedings of the 43rd annual ACM
  symposium on Theory of computing}}, \bibinfo{pages}{333--342}
  (\bibinfo{organization}{ACM}, \bibinfo{year}{2011}).

\bibitem{Broome2013}
\bibinfo{author}{Broome, M.~A.} \emph{et~al.}
\newblock \bibinfo{title}{Photonic boson sampling in a tunable circuit}.
\newblock \emph{\bibinfo{journal}{Science}} \textbf{\bibinfo{volume}{339}},
  \bibinfo{pages}{794--798} (\bibinfo{year}{2013}).

\bibitem{Spring2013}
\bibinfo{author}{Spring, J.~B.} \emph{et~al.}
\newblock \bibinfo{title}{Boson sampling on a photonic chip}.
\newblock \emph{\bibinfo{journal}{Science}} \textbf{\bibinfo{volume}{339}},
  \bibinfo{pages}{798--801} (\bibinfo{year}{2013}).

\bibitem{Tillmann2013}
\bibinfo{author}{Tillmann, M.} \emph{et~al.}
\newblock \bibinfo{title}{Experimental boson sampling}.
\newblock \emph{\bibinfo{journal}{Nature Photonics}}
  \textbf{\bibinfo{volume}{7}}, \bibinfo{pages}{540--544}
  (\bibinfo{year}{2013}).

\bibitem{Crespi2013}
\bibinfo{author}{Crespi, A.} \emph{et~al.}
\newblock \bibinfo{title}{Integrated multimode interferometers with arbitrary
  designs for photonic boson sampling}.
\newblock \emph{\bibinfo{journal}{Nature Photonics}}  (\bibinfo{year}{2013}).

\bibitem{Reck1994}
\bibinfo{author}{Reck, M.}, \bibinfo{author}{Zeilinger, A.},
  \bibinfo{author}{Bernstein, H.~J.} \& \bibinfo{author}{Bertani, P.}
\newblock \bibinfo{title}{Experimental realization of any discrete unitary
  operator}.
\newblock \emph{\bibinfo{journal}{Physical Review Letters}}
  \textbf{\bibinfo{volume}{73}}, \bibinfo{pages}{58} (\bibinfo{year}{1994}).

\bibitem{Hong1987}
\bibinfo{author}{Hong, C.}, \bibinfo{author}{Ou, Z.} \&
  \bibinfo{author}{Mandel, L.}
\newblock \bibinfo{title}{Measurement of subpicosecond time intervals between
  two photons by interference}.
\newblock \emph{\bibinfo{journal}{Physical Review Letters}}
  \textbf{\bibinfo{volume}{59}}, \bibinfo{pages}{2044--2046}
  (\bibinfo{year}{1987}).

\bibitem{Dhand2016}
\bibinfo{author}{Dhand, I.}, \bibinfo{author}{Khalid, A.}, \bibinfo{author}{Lu,
  H.} \& \bibinfo{author}{Sanders, B.~C.}
\newblock \bibinfo{title}{Accurate and precise characterization of linear
  optical interferometers}.
\newblock \emph{\bibinfo{journal}{Journal of Optics}}
  \textbf{\bibinfo{volume}{18}}, \bibinfo{pages}{035204}
  (\bibinfo{year}{2016}).

\bibitem{Tillmann2015}
\bibinfo{author}{Tillmann, M.} \emph{et~al.}
\newblock \bibinfo{title}{Generalized multiphoton quantum interference}.
\newblock \emph{\bibinfo{journal}{Phys. Rev. X}} \textbf{\bibinfo{volume}{5}},
  \bibinfo{pages}{041015} (\bibinfo{year}{2015}).

\bibitem{Peruzzo2011}
\bibinfo{author}{Peruzzo, A.}, \bibinfo{author}{Laing, A.},
  \bibinfo{author}{Politi, A.}, \bibinfo{author}{Rudolph, T.} \&
  \bibinfo{author}{O'Brien, J.~L.}
\newblock \bibinfo{title}{Multimode quantum interference of photons in
  multiport integrated devices}.
\newblock \emph{\bibinfo{journal}{Nature communications}}
  \textbf{\bibinfo{volume}{2}}, \bibinfo{pages}{224} (\bibinfo{year}{2011}).

\bibitem{Edelman1988}
\bibinfo{author}{Edelman, A.}
\newblock \bibinfo{title}{Eigenvalues and condition numbers of random
  matrices}.
\newblock \emph{\bibinfo{journal}{SIAM Journal on Matrix Analysis and
  Applications}} \textbf{\bibinfo{volume}{9}}, \bibinfo{pages}{543--560}
  (\bibinfo{year}{1988}).

\bibitem{Peres1989}
\bibinfo{author}{Peres, A.}
\newblock \bibinfo{title}{Construction of unitary matrices from observable
  transition probabilities}.
\newblock \emph{\bibinfo{journal}{Nuclear Physics B-Proceedings Supplements}}
  \textbf{\bibinfo{volume}{6}}, \bibinfo{pages}{243--245}
  (\bibinfo{year}{1989}).

\bibitem{taylor1997introduction}
\bibinfo{author}{Taylor, J.}
\newblock \bibinfo{title}{Introduction to error analysis, the study of
  uncertainties in physical measurements}.
\newblock \emph{\bibinfo{journal}{Published by University Science Books, 648
  Broadway, Suite 902, New York, NY 10012, 1997.}} \textbf{\bibinfo{volume}{1}}
  (\bibinfo{year}{1997}).

\bibitem{Peruzzo2010}
\bibinfo{author}{Peruzzo, A.} \emph{et~al.}
\newblock \bibinfo{title}{Quantum walks of correlated photons}.
\newblock \emph{\bibinfo{journal}{Science}} \textbf{\bibinfo{volume}{329}},
  \bibinfo{pages}{1500--1503} (\bibinfo{year}{2010}).

\bibitem{Sansoni2012}
\bibinfo{author}{Sansoni, L.} \emph{et~al.}
\newblock \bibinfo{title}{Two-particle bosonic-fermionic quantum walk via
  integrated photonics}.
\newblock \emph{\bibinfo{journal}{Physical review letters}}
  \textbf{\bibinfo{volume}{108}}, \bibinfo{pages}{010502}
  (\bibinfo{year}{2012}).

\bibitem{Biggerstaff2016}
\bibinfo{author}{Biggerstaff, D.~N.} \emph{et~al.}
\newblock \bibinfo{title}{Enhancing coherent transport in a photonic network
  using controllable decoherence}.
\newblock \emph{\bibinfo{journal}{Nature communications}}
  \textbf{\bibinfo{volume}{7}} (\bibinfo{year}{2016}).

\bibitem{Musha1982}
\bibinfo{author}{Musha, T.}, \bibinfo{author}{Kamimura, J.-i.} \&
  \bibinfo{author}{Nakazawa, M.}
\newblock \bibinfo{title}{Optical phase fluctuations thermally induced in a
  single-mode optical fiber}.
\newblock \emph{\bibinfo{journal}{Applied optics}}
  \textbf{\bibinfo{volume}{21}}, \bibinfo{pages}{694--698}
  (\bibinfo{year}{1982}).

\bibitem{hati2011vibration}
\bibinfo{author}{Hati, A.}, \bibinfo{author}{Nelson, C.} \&
  \bibinfo{author}{Howe, D.}
\newblock \bibinfo{title}{Vibration sensitivity of optical components: a
  survey}.
\newblock In \emph{\bibinfo{booktitle}{Frequency Control and the European
  Frequency and Time Forum (FCS), 2011 Joint Conference of the IEEE
  International}}, \bibinfo{pages}{1--4} (\bibinfo{organization}{IEEE},
  \bibinfo{year}{2011}).

\bibitem{Jin2014}
\bibinfo{author}{Jin, R.-B.} \emph{et~al.}
\newblock \bibinfo{title}{Pulsed sagnac polarization-entangled photon source
  with a ppktp crystal at telecom wavelength}.
\newblock In \emph{\bibinfo{booktitle}{CLEO: QELS\_Fundamental Science}},
  \bibinfo{pages}{FF1D--3} (\bibinfo{organization}{Optical Society of America},
  \bibinfo{year}{2014}).

\bibitem{Bentivegna2015}
\bibinfo{author}{Bentivegna, M.} \emph{et~al.}
\newblock \bibinfo{title}{Experimental scattershot boson sampling}.
\newblock \emph{\bibinfo{journal}{Science Advances}}
  \textbf{\bibinfo{volume}{1}}, \bibinfo{pages}{e1400255}
  (\bibinfo{year}{2015}).

\bibitem{Burr1942}
\bibinfo{author}{Burr, I.~W.}
\newblock \bibinfo{title}{Cumulative frequency functions}.
\newblock \emph{\bibinfo{journal}{The Annals of Mathematical Statistics}}
  \textbf{\bibinfo{volume}{13}}, \bibinfo{pages}{215--232}
  (\bibinfo{year}{1942}).

\end{thebibliography}
%


\end{document}